\documentclass[twocolumn,english,aps,prl,longbibliography,showpacs,amsfonts,amsmath,amssymb,superscriptaddress,footinbib,floatfix]{revtex4}
\usepackage[T1]{fontenc}
\usepackage[latin9]{inputenc}
\usepackage{amstext}
\usepackage{amssymb}
\usepackage{amsmath}
\usepackage{graphicx}
\usepackage{subfigure}
\usepackage{ulem}
\usepackage[dvipsnames]{xcolor}

\usepackage{filemod}

\def\kbar{{\mathchar'26\mkern-9mu k}}

\begin{document}

\title{Experimental observation of time singularity in classical-to-quantum chaos transition}

\author{Cl{\' e}ment Hainaut}
\affiliation{Universit{\'e} de Lille, CNRS, UMR 8523 -- PhLAM -- Laboratoire de Physique des Lasers Atomes et Mol{\'e}cules, F-59000 Lille, France}
\homepage{www.phlam.univ-lille1.fr/atfr/cq}

\author{Ping Fang}
\affiliation{Institute for Advanced Study, Tsinghua University, Beijing 100084,China}
\affiliation{CAS Key Laboratory of Frontiers in Theoretical Physics and Institute of Theoretical Physics, Chinese Academy of Sciences, Beijing 100190, China}

\author{Adam Ran{\c c}on}
\affiliation{Universit{\'e} de Lille, CNRS, UMR 8523 -- PhLAM -- Laboratoire de Physique des Lasers Atomes et Mol{\'e}cules, F-59000 Lille, France}
\homepage{www.phlam.univ-lille1.fr/atfr/cq}

\author{Jean-Fran{\c c}ois Cl{\' e}ment}
\affiliation{Universit{\'e} de Lille, CNRS, UMR 8523 -- PhLAM -- Laboratoire de Physique des Lasers Atomes et Mol{\'e}cules, F-59000 Lille, France}
\homepage{www.phlam.univ-lille1.fr/atfr/cq}

\author{Pascal Szriftgiser}
\affiliation{Universit{\'e} de Lille, CNRS, UMR 8523 -- PhLAM -- Laboratoire de Physique des Lasers Atomes et Mol{\'e}cules, F-59000 Lille, France}
\homepage{www.phlam.univ-lille1.fr/atfr/cq}

\author{Jean-Claude Garreau}
\affiliation{Universit{\'e} de Lille, CNRS, UMR 8523 -- PhLAM -- Laboratoire de Physique des Lasers Atomes et Mol{\'e}cules, F-59000 Lille, France}
\homepage{www.phlam.univ-lille1.fr/atfr/cq}

\author{Chushun Tian}
\email{ct@itp.ac.cn}
\affiliation{CAS Key Laboratory of Frontiers in Theoretical Physics and Institute of Theoretical Physics, Chinese Academy of Sciences, Beijing 100190, China}
\affiliation{Universit{\'e} de Lille, CNRS, UMR 8523 -- PhLAM -- Laboratoire de Physique des Lasers Atomes et Mol{\'e}cules, F-59000 Lille, France}
\homepage{www.phlam.univ-lille1.fr/atfr/cq}

\author{Radu Chicireanu}
\email{rchicireanu@gmail.com}
\affiliation{Universit{\'e} de Lille, CNRS, UMR 8523 -- PhLAM -- Laboratoire de Physique des Lasers Atomes et Mol{\'e}cules, F-59000 Lille, France}
\affiliation{CAS Key Laboratory of Frontiers in Theoretical Physics and Institute of Theoretical Physics, Chinese Academy of Sciences, Beijing 100190, China}
\homepage{www.phlam.univ-lille1.fr/atfr/cq}

\date{\today}

\begin{abstract}
The emergence of chaotic phenomena in a quantum system has long been an elusive subject. Experimental progresses in this subject have become urgently needed in recent years, when considerable theoretical studies have unveiled the vital roles of chaos in a broad range of topics in quantum physics. Here, we report the first experimental observation of time singularity, that signals a classical-to-quantum chaos transition and finds its origin in the {\it sudden change} in system's memory behaviors. The time singularity observed is an analog of the {\it dynamical quantum phase transition} (DQPT) -- proposed very recently for regular systems -- in chaotic systems, but with totally different physical origin.

\end{abstract}


\maketitle

Classical and quantum chaos are two sides of the same coin, but behave in a very different, even opposite, way. Notably, in the former dynamical instabilities erase the memory of initial states, whereas in the latter the dynamics is stable, and this memory is kept during the time evolution of the system. A natural question for quantized classical chaotic systems is how quantum effects and classical chaos interplay and give rise to this memory recovery, and the ensuing emergence of quantum chaos. This is important not only to the fundamental subject of the quantum-classical correspondence in chaotic systems, but also for realizing quantum control of chaos. Furthermore, it may shed new light on various intriguing quantum chaotic phenomena found in different fields recently (e.g., Refs.~\cite{Preskill07:JHEP,Kitaev15,Kitaev15:KITP,Maldacena16:PRD,Maldacena16:JHEP,Sachdev17:PNAS,
Ioffe16:Ann.Phys.,Chen:QuantumDrivenIntegerQuantumHallEffect:PRL14,Tian:EmergencerQuantumHallEffectChaos:PRB16}). Notwithstanding that the interplay between quantum effects and classical chaos have been investigated from distinct aspects theoretically (e.g.~Refs.~\cite{Shepelyansky87:PhysicaD,CasatiChirikov95,Larkin96:PRB,Sokolov08:EPL,Casati09:PRE}), most experiments focus on the deep quantum regime, where information on classical chaos is difficult to extract, and thus tell us nothing about this interplay.

\begin{figure}[h]
\includegraphics[width=8.3cm] {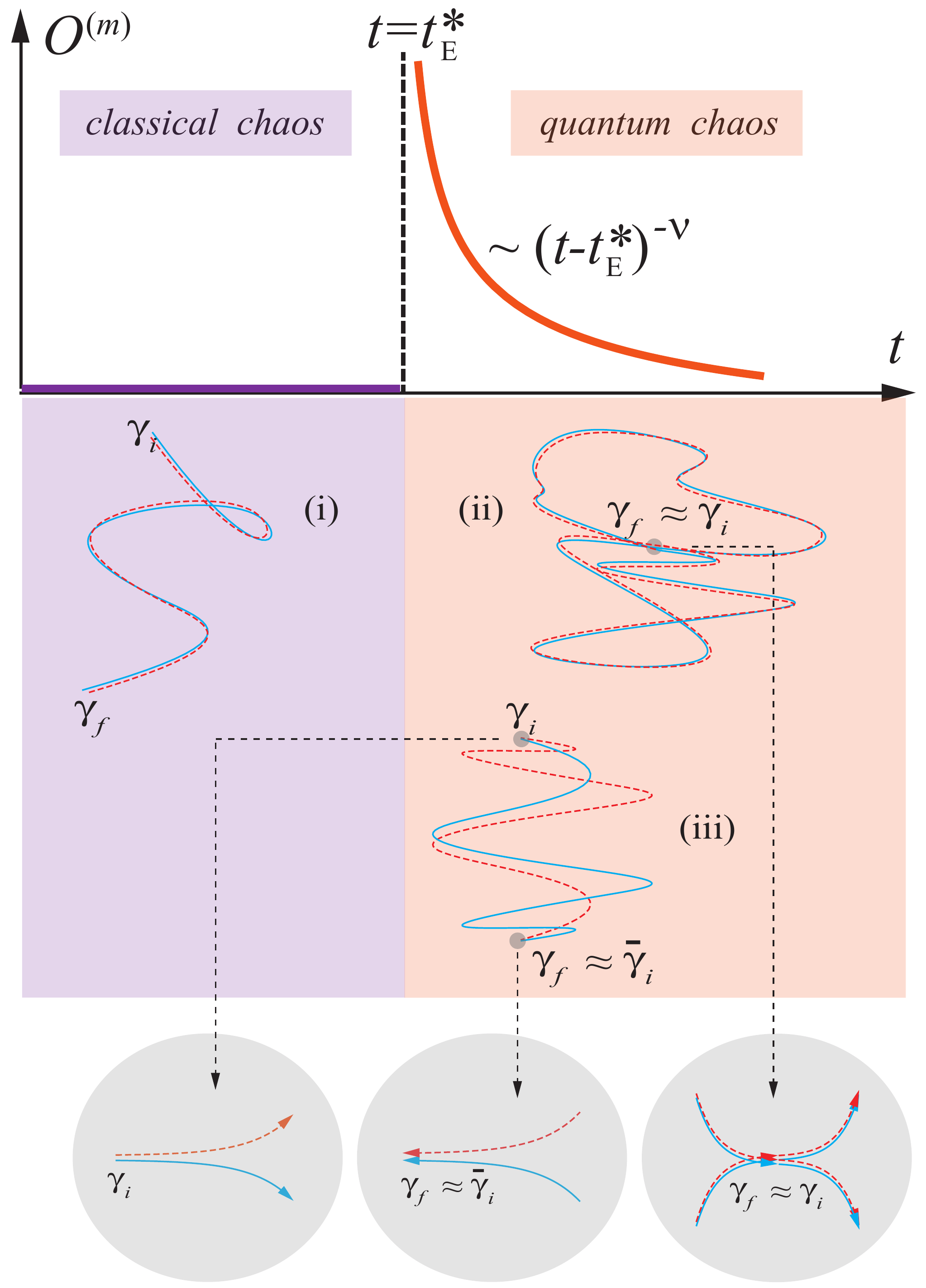}
\caption{Physical mechanism for time singularity. At the critical time $t_{\rm E}^*$ (top) the system's memory behavior undergoes a sudden change (middle). For $t<t_{\rm E}^*$ the system wanders randomly in phase space and the initial and final states, $\gamma_i$ and $\gamma_f$, are uncorrelated (i). For $t>t_{\rm E}^*$ two trajectories (blue solid and red dashed lines) departing from Planck's cell at $\gamma_i$, due to dynamical instability (bottom), can interfere constructively and meet again with a significant probability in the same cell (ii) or the cell (iii) at ${\bar \gamma}_i$ (which is the time reversed conjugate of $\gamma_i$) after long wandering; thus the memory of $\gamma_i$ is recovered.}
\label{fig:3}
\end{figure}

Here we explore this subject experimentally by using the atom-optics realization of a canonical model in nonlinear dynamics, the quantum kicked rotor (QKR) \cite{Casati:LocDynFirst:LNP79}, and its variants. We measure the time evolution of certain observable (to be defined below) $O(t)$ and find that, as shown schematically in Fig.~\ref{fig:3}, it displays a singularity at a critical time, namely, the Ehrenfest time $t_{\rm E}^*$ \cite{Larkin:Ehrenfest_time:JETP68}. The Ehrenfest time results from the amplification of Planck's cell by dynamical instability, and signals the breakdown of quantum-classical correspondence. At time $t=t_{\rm E}^*$, the behavior of $O(t)$ undergoes a {\it sudden change},
\begin{eqnarray}
O^{(m)}(t)=\,\Big\{\begin{array}{c}
                              0,\quad\quad t<t_{\rm E}^*;\qquad\qquad \\
                              \!\!\!\!c (t-t_{\rm E}^*)^{-\nu},\quad t>t_{\rm E}^*,
                            \end{array}
\label{eq:5}
\end{eqnarray}
where the superscript $(m)$ stands for the $m(\in \mathbb{N})$th order derivative, and $\nu\geq 0$. Moreover, the lower order derivatives of $O(t)$ are regular. The values of $m,\nu$ are determined by system's symmetry, but not by the detailed construction of QKR. The latter only affects the values of $t_{\rm E}^*$ and the proportionality coefficient $c$. We argue that this dynamical phenomenon is analogous to the continuous phase transitions in statistical mechanics: $t$ mimics the (inverse) temperature, $O(t)$ the order parameter, and $\nu$ the critical exponent. This singularity was predicted analytically in Refs.~\cite{Tian:EhrenfestTimeDynamicalLoc:PRL04,Tian:EhrenfestTimeDynamicalLoc:PRB05}, and has been observed in a numerical study of the out-of-time-order correlator \cite{Galitskii:out_of_time_order_correlator}. We further show in the supplemental material \cite{SM} that it is intimately related to a nonanalytic behavior of the return probability at the critical time. Thus the time singularity observed is analogous to the DQPT referred to nonanalytic behaviors of the return probability at a critical time, which was predicted originally for regular systems and has been confirmed experimentally~\cite{Heyl:DynamicalPhaseTransition:PRL13,Jurcevic:DynamicalPhaseTransition:PRL17,Heyl:DynamicalPhaseTransition:arXiv17}. As discussed below, the singularity finds its origin in a sudden change in system's memory behavior. Thus it is a sharp border between classical and quantum chaos in the time domain. 

To see the origin mentioned above (cf.~Fig.~\ref{fig:3}) we work in the Wigner representation, which yields a ``natural'' connection between classical and quantum mechanics. In this representation a quantum state corresponds to a Planck's cell in the phase space. For short evolution time, $t<t_{\rm E}^*$, the cell's center moves along a classical trajectory -- the quantum-classical correspondence -- and fast separation between nearby trajectories (i.e., dynamical instability) renders the memory of the initial state $\gamma_i$ lost in the course of time. As a result, the final state $\gamma_f$ is uncorrelated with $\gamma_i$, and the system wanders randomly in phase space. For long evolution time, $t>t_{\rm E}^*$, trajectories departing from the same Planck's cell reach a large separation at $t=t_{\rm E}^*$, and wander independently at later times. However, these trajectories have a chance to meet again in the initial cell $\gamma_i$ or its time reversed ${\bar \gamma}_i$, depending on whether the time-reversal ($T_c$) symmetry is present. In particular, when two such trajectories in phase space are piecewise identical or identical up to the time reversal, their quantum amplitudes have the same phases and thus they can interfere constructively. Consequently, the probability for ``meeting'' in the vicinity of $\gamma_i$ or ${\bar \gamma}_i$ is enhanced. This enhancement has recently been observed experimentally in the QKR~\cite{Hainaut:ERO:PRL17}. As such, the memory of $\gamma_i$ is recovered, and we see that both the dynamical instability and quantum interference are indispensable for this recovery. This physical picture is quite general. In particular, it has nothing to do with the system's eventual fate [e.g., (de)localization], and is independent of the choice of observables. Taking this degree of freedom, throughout this work we choose $O(t)$ to be what is defined by Eq.~(\ref{eq:1}) below. It can be shown \cite{SM} that the corresponding $O^{(2)}(t)$ gives the time correlation of the angular position of QKR.

Our experimental realization of the QKR consists of an atom of mass $M$ which is submitted to a series of short pulses (kicks) of a one-dimensional sinusoidal potential applied periodically in time, at intervals $T_1$, during very short periods of time $\tau$. The potential is realized using a far-detuned laser standing wave (SW), formed by two independent counter-propagating laser beams, with wavenumber $k_{L}$. For $\tau\ll T_1$ the pulse can be considered as a Dirac function, and thus the Hamiltonian describing the atom's motion is:
\begin{equation}
\hat H(t)=\frac{\hat p^{2}}{2}+K\cos\left(\hat x-a(t)\right)\sum_{n}\delta(t-n).\label{eq:Ha}
\end{equation}
Here, the time unit is $T_1$, the space unit is $1/(2k_{L})$, and the momentum unit is $M/(2k_{L}T_1)$. In these units, $[\hat x,\hat p]=i\kbar$, where $\kbar=4\hbar k_{L}^{2}T_{1}/M$ is reduced Planck's constant and $\hbar$ is the Planck's constant. The normalized strength $K$ of the kicks, called the stochasticity
parameter, is $\propto I/|\Delta|$, where $I$ is the maximum laser intensity and $\Delta$ the laser-atom detuning.

We load a cloud of about $10^{6}$ cesium atoms in a standard magneto-optical trap and cool it to a temperature of few microkelvins by an optimized molasses phase. The cloud is then exposed to a pulsed, vertical SW with the following parameters: laser beam wavelength $852$ nm (D$2$ line), $T_1=9.6$ $\mu{\rm s}$, $\tau$ in the range of $200-300$ ${\rm ns}$, $\Delta\approx-13$ GHz (so that spontaneous emission can be neglected for the duration of experiments), waist $0.8$ mm and $I\approx30$ W/cm$^2$, corresponding to $K$ in the range of $4-12$. We estimate an about $5$-$10 \%$ inhomogeneity in $K$ due to the finite transverse extension ($150-250$ $\mu$m) of the atomic cloud. By adding a linear chirp of the relative frequency of the beams, we generate an SW whose nodes are accelerated, and this acceleration is adjusted to be exactly equal to the gravity's acceleration. Hence in the (non-inertial) reference frame in which the SW is at rest, an inertial force exactly compensates the effect of gravity \cite{Manai:Anderson2DKR:PRL15}. At the end of the kick sequence, the atomic momentum distribution $\Pi(p,t)$ is detected by a standard time-of-flight measurement of a duration of $170 {\rm ms}$. From $\Pi(p,t)$ we obtain the cloud expansion~\cite{Note1}:
\begin{equation}\label{eq:1}
    \delta\langle p^2(t) \rangle=\int dpp^2 (\Pi(p,t)-\Pi(p,0))\equiv O(t),
\end{equation}
which defines the aforementioned observable to be studied in this work. In order to explore the transition from classical to quantum chaos a small $\kbar$ is desired, so that the time scale at which localization effects \cite{Casati:LocDynFirst:LNP79,Fishman:LocDynAnders:PRL82} dominate can be long enough. However, because the requirement: $T_1\gg \tau$ introduces a limitation on the minimal value of $\kbar$, only $\kbar \geq 1$ can be achieved in our experiments~\cite{Note2}.

The phase shift $a(t)$ in Eq.~\eqref{eq:Ha} is the relative phase between the two laser beams, and its profile can be readily changed.
When $a$ is a constant, the model described by Eq.~(\ref{eq:Ha}) reduces to the standard QKR \cite{Casati:LocDynFirst:LNP79}. When $a(t)$ is modulated by $(d-1)$ incommensurate frequencies, which are all incommensurate with $2\pi$, a quasiperiodic QKR results, which is equivalent to a $d$-dimensional periodic QKR \cite{Casati:IncommFreqsQKR:PRL89}. Investigations into effects of periodic modulation, i.e., $a(t + N)=a(t)$ with $N$ being an integer, were initiated in Ref.~\cite{Tian:EhrenfestTimeDynamicalLoc:PRB05}; substantial progresses have been made recently \cite{Manai:Anderson2DKR:PRL15,Hainaut:CFS:arXiv17}. This variant of the standard QKR is called the periodically-shifted QKR~\cite{Hainaut:CFS:arXiv17}. Thus $\hat H(t)$ describes a generalized QKR system.

Depending on the time dependence of $a(t)$, this generalized QKR system exhibits very different dynamical behaviors. In particular, the periodic modulation allows one to explore rich symmetry effects. Indeed, the standard QKR ($N$$=$$1$) possesses the (effective) $T_c$ symmetry. Namely, $\hat H(t)$ is invariant under the transformation: $t \rightarrow -t$, $\hat x \rightarrow -\hat x$, $\hat p \rightarrow \hat p$. Note that $\hat x$ mimics the electron momentum and $\hat p$ the position in conventional disordered electronic systems \cite{Efetov:SupersymmetryInDisorder:97}. Equivalently, the Floquet operator: $\hat U\equiv e^{-\frac{i}{4\kbar}\hat p^2}e^{-\frac{iK}{\kbar}\cos\hat x}e^{-\frac{i}{4\kbar}\hat p^2}$, governing the dynamics at integer times, is invariant under the matrix transposition (up to unitary transformations corresponding to space and time translations),
\begin{equation}\label{eq:T_symmetry}
    (\hat U)_{pp'}=(\hat U)_{p'p}\,\;\; \Rightarrow\,\;\; T_c\,\;\; {\rm symmetry}.
\end{equation}
As a result of this symmetry, the dynamical localization in standard QKR \cite{Casati:LocDynFirst:LNP79,Fishman:LocDynAnders:PRL82} is found to be an analog of Anderson localization in quasi one-dimensional disordered systems \cite{Tian:TheoryOfLocalizationQKR:NJP10} in the orthogonal class (in jargon of the random matrix theory). More universality classes can be realized in periodically-shifted QKR ($N>2$) \cite{Tian:EhrenfestTimeDynamicalLoc:PRB05,Hainaut:CFS:arXiv17}. In particular, as adopted below, one can randomly choose the modulation configuration: $\{a(0)=0,a(1),\cdots, a(N-1)\}$. In this case, the Floquet operator is a product of successive $N$ one-step evolution operators, and can be checked to break the $T_c$ symmetry (\ref{eq:T_symmetry})~\cite{Note3}.
Dynamical localization is then analogous to that of Anderson localization in the unitary class \cite{Tian:EhrenfestTimeDynamicalLoc:PRB05}. The symmetry effects of periodic modulation can be seen readily from fluctuations of the quasieigenenergy spectrum of Floquet operators~\cite{SM}.

We now present experimental results and their theoretical analysis, starting with the standard QKR ($N$$=$$1$) which exhibits the $T_c$ symmetry. Since the decoherence rate, which introduces a limitation on the duration of experiments, increases with $K$, we choose $K=5.5$. For this $K$ the classical dynamics is known to be strongly chaotic \cite{Chirikov:ChaosClassKR:PhysRep79}. We choose $\kbar=1$.

\begin{figure}[t!]
\includegraphics[width=8.3cm] {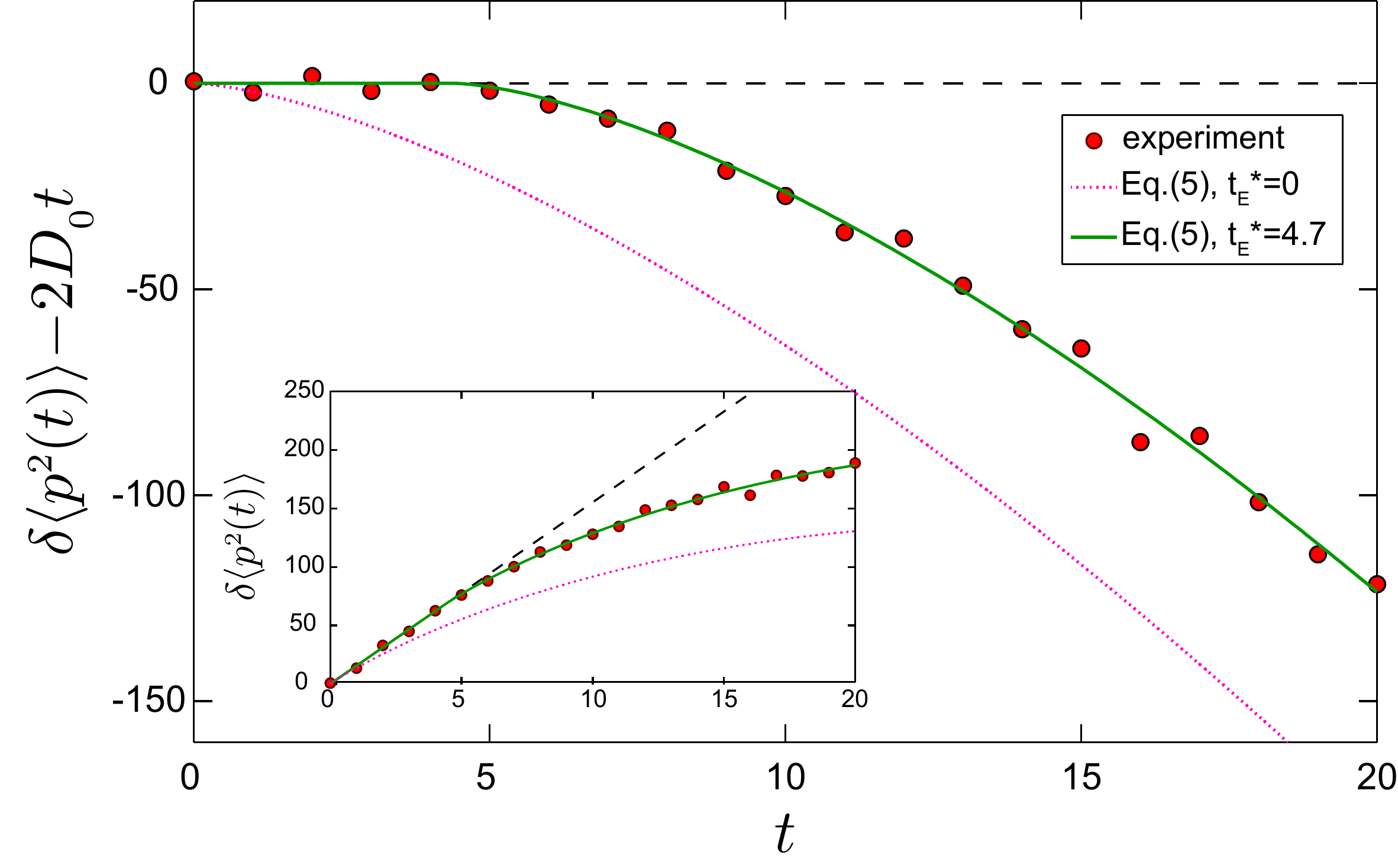}
\caption{Experiments on the standard QKR ($K=5.5,\kbar=1$), that displays $T_c$ symmetry, confirm a classical-to-quantum chaos transition, manifesting in the time profile of $\delta \langle p^2(t) \rangle-2D_0t$. Experimental data (symbols) is very well described by Eq.~(\ref{eq:4}) with $t_{\rm loc}$ and $t_{\rm E}^*$ treated as fitting parameters, and show a time singularity at $t_{\rm E}^*=4.7$ (green solid line). The data is incompatible with $t_{\rm E}^*=0$ (pink dotted line). Inset: Same as main panel, showing  $\delta \langle p^2(t) \rangle$ instead. The black dashed line represents the diffusion at early times.}
\label{fig:1}
\end{figure}

The measurements of $\delta \langle p^2(t)\rangle$ are displayed in the inset of Fig.~\ref{fig:1}. At early times, its growth is linear, $\delta \langle p^2(t) \rangle=2D_0t$ with $D_0$ the diffusion coefficient, corresponding to the classical chaotic behavior. This  implies that the system undergoes random wandering in phase space [(i) in Fig.~\ref{fig:3}] and exhibits a (normal) diffusion in rotor's momentum ($p$) direction. To better analyze the behaviors at later times, we also show in Fig.~\ref{fig:1} $\delta \langle p^2(t) \rangle$$-$$2D_0t$. We clearly observe that the growth is purely linear up to a critical time $t_E^*$, from which it starts to deviate from the classical behaviors.

To determine $t_{\rm E}^*$ and study the dynamical behaviors beyond this critical time, we use the Ehrenfest time-dependent weak dynamical localization theory \cite{Tian:EhrenfestTimeDynamicalLoc:PRB05,Tian:EhrenfestTimeDynamicalLoc:PRL04}, which gives for the standard QKR:
\begin{equation}\label{eq:4}
    \delta \langle p^2(t) \rangle=2D_0\left(t -\frac{4}{3\sqrt{\pi}}\theta(t-t_{\rm E}^*)\frac{(t-t_{\rm E}^*)^{3/2}}{t_{\rm loc}^{1/2}}\right)
\end{equation}
with $t_{\rm loc}$ the localization time and $\theta(t)$ the Heavisde function. Equation (\ref{eq:4}) shows that $\delta \langle p^2(t)\rangle$ measured at integer $t$ can be embedded into a curve which is continuous in $t$, and this curve exhibits a singularity (cf.~Fig.~\ref{fig:3}) at the Ehrenfest time $t_{\rm E}^*$ (not necessarily an integer), with:
\begin{equation}\label{eq:2}
    m=2,\quad \nu=1/2.
\end{equation}
The scaling law $\sim (t-t_{\rm E}^*)^{3/2}$ accounts for the weak dynamical localization, arising from constructive interference between a trajectory and its time reversal [(iii) in Fig.~\ref{fig:3}] which exists for $t\geq t_{\rm E}^*$. This formula was derived analytically for $K$$\gg $$1$$\gg$$ \kbar$. For present parameters: $K$$=$$5.5$, $\kbar$$=$$1$, we expect that it remains valid qualitatively -- as the physical picture of the second term is quite general, but not quantitatively. We will thus use $t_{\rm loc}$ as an additional fitting parameter. Our fitting procedure of the experimental data is as follows: First of all, we pick a time $t_0$, and perform a {\em linear} fit $ \delta \langle p^2(t) \rangle$ from $t=0$ to $t=t_0$ to extract a diffusion coefficient
$D_0$; we then perform a two-parameter fit of the full data using $t_{\rm E}^*$ and $t_{\rm loc}$ as parameters (keeping $D_0$ fixed); if $t_0$$>$$t_{\rm E}^*$ or if  $t_0\ll t_{\rm E}^*$, we re-estimate $D_0$ by setting $t_0$ to $t_{\rm E}^*$ and repeat the procedure until $t_0$ and $t_{\rm E}^*$ converge.
As shown in Fig.~\ref{fig:1}, the measurements are in excellent agreement with Eq.~(\ref{eq:4}) for $t_{\rm loc}=33$ and $t_{\rm E}^*=4.7$. A vanishing $t_{\rm E}^*$, as predicted in the QKR analog~\cite{Altland:DiagrammaticAndersonLocQKR:PRL93} of standard weak localization in disordered systems \cite{Gorkov:WeakLocalization:JETP79} is clearly inconsistent with the experimental data.

\begin{figure}[t!]
\includegraphics[width=8.3cm]{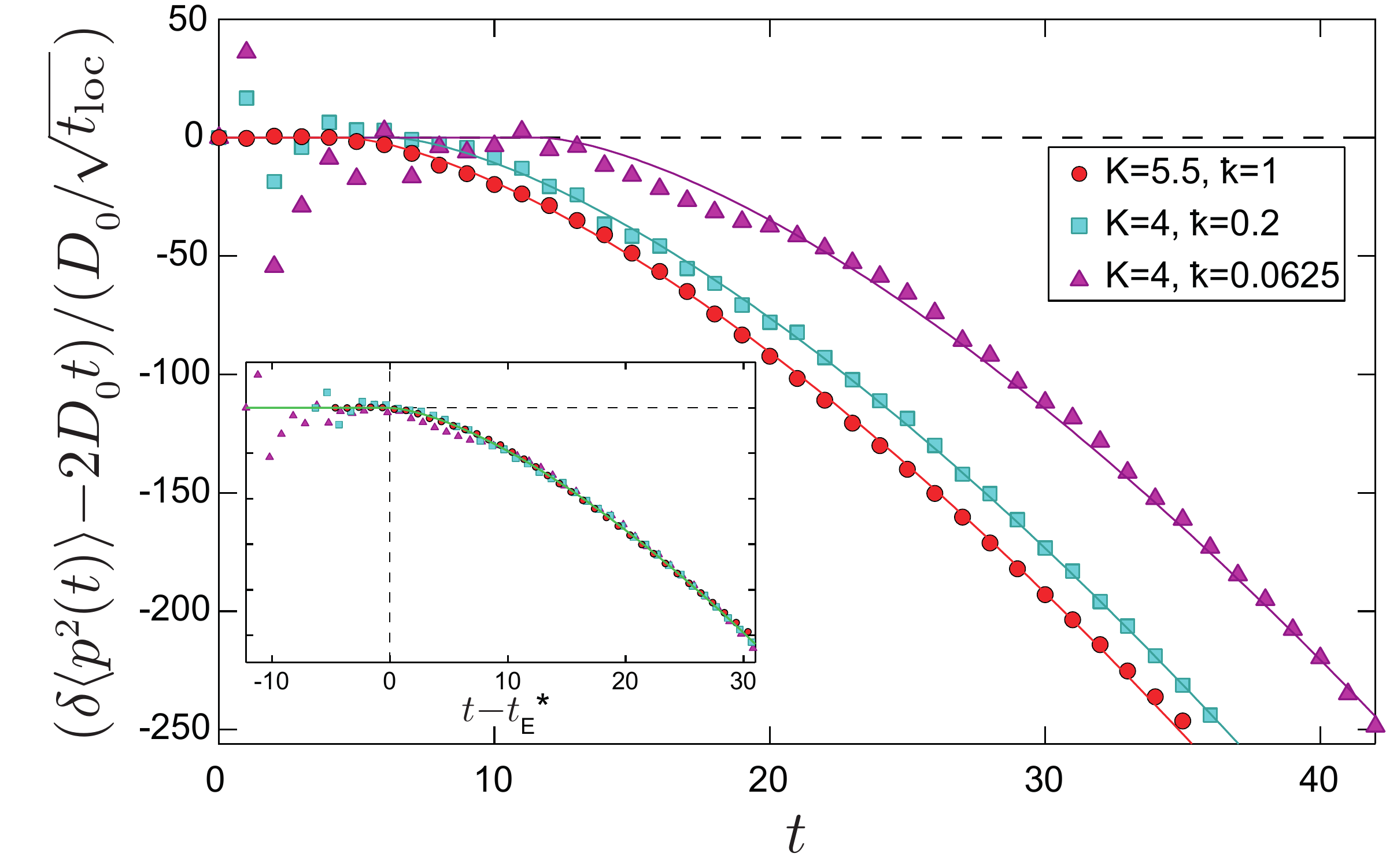}
\caption{Results of numerical simulations for different values of $\kbar<1$ (symbols), corresponding to the weak dynamical localization correction term in Eq.~\eqref{eq:4}, scaled by $D_0/\sqrt{t_{\rm loc}}$. The results are very well described by Eq.~\eqref{eq:4} (solid lines). The time singularity occurs at $t_{\rm E}^*=4.7$ ($t_{\rm loc}=39$) for $\kbar=1$,	$t_{\rm E}^*=6.3$ ($t_{\rm loc}=172$) for $\kbar=0.2$ and $t_{\rm E}^*=11.8$ ($t_{\rm loc}=1200$) for $\kbar=0.0625$. Inset: Same as the main panel, with the time axis shifted by $-t_{\rm E}^*$. Beyond $t_{\rm E}^*$, all data collapse into a universal curve, in very good agreement with the $\sim (t-t_{\rm E}^*)^{3/2}$ scaling law (solid line).}
\label{fig:4}
\end{figure}

To better understand the physical meaning of $t_{\rm E}^*$ we further perform numerical simulations for values of $\kbar<1$ which, as discussed above, cannot be achieved in our experiments. The results of $\delta \langle p^2(t)\rangle$ for different $\kbar$ are displayed Fig.~\ref{fig:4}. They are in excellent agreement with Eq.~(\ref{eq:4}). (We use the same fitting procedure as above to extract first $D_0$ and then $t_{\rm loc}$ and $t_{\rm E}^*$.) Most importantly, we see that, when keeping $K$ approximately constant, the critical time increases as $\kbar$ decreases (main panel). This is consistent with the physical interpretation that the Ehrenfest time signals the breakdown of quantum-classical correspondence and thus should increases with decreasing $\kbar$ \cite{Larkin96:PRB,Sokolov08:EPL,Casati09:PRE,Larkin:Ehrenfest_time:JETP68,Tian:EhrenfestTimeDynamicalLoc:PRL04,Tian:EhrenfestTimeDynamicalLoc:PRB05,Galitskii:out_of_time_order_correlator}, although we do not have analytical formula for this regime of $K,\kbar$. In addition, when we shift $t$ by $-t_{\rm E}^*$ and rescale $\delta \langle p^2(t) \rangle$$-$$2D_0t$, all data collapse into a universal curve corresponding to the scaling law $\sim (t-t_{\rm E}^*)^{3/2}$ (inset).

\begin{figure}[t!]
\includegraphics[width=8.3cm] {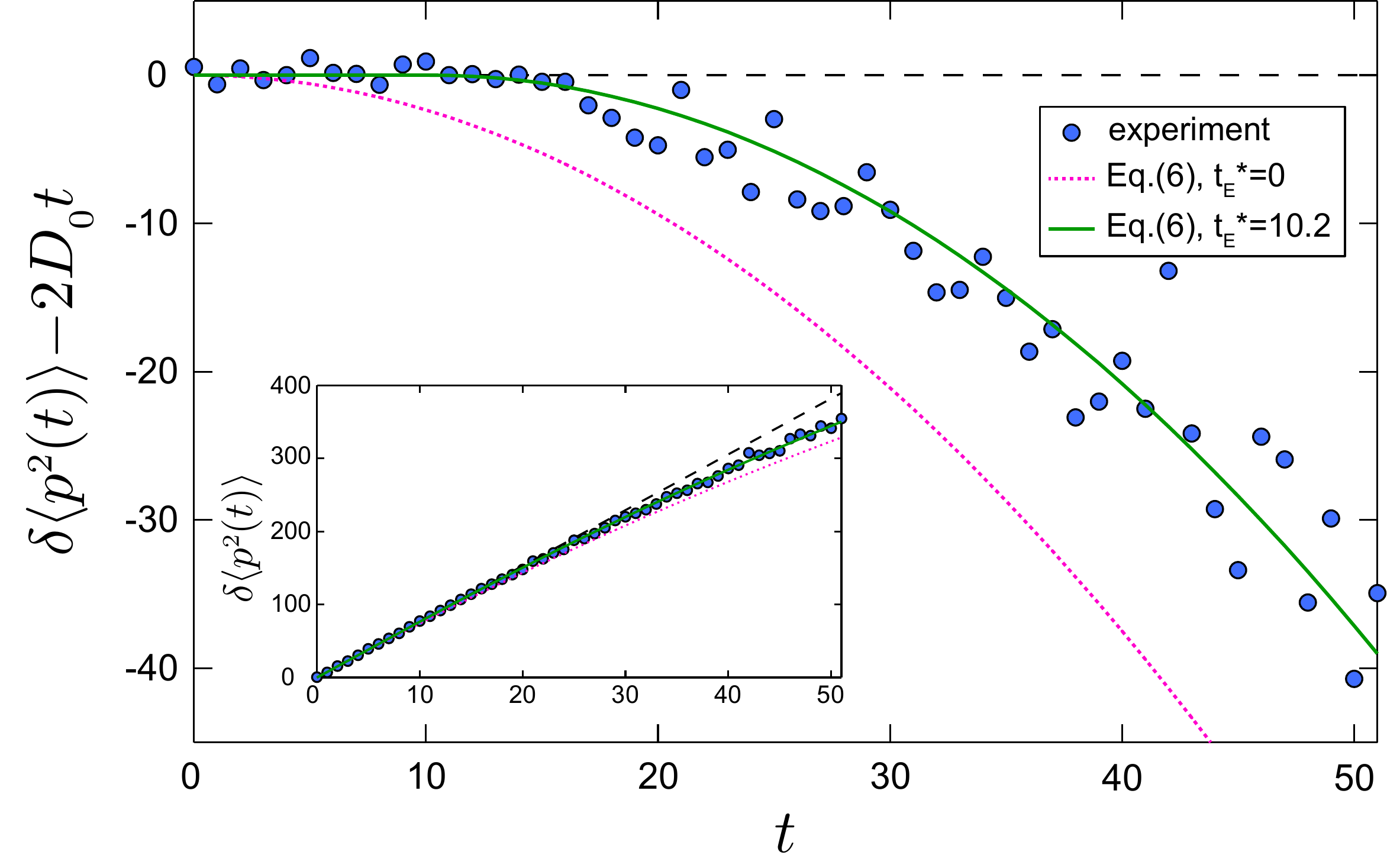}
\caption{Classical-to-quantum chaos transition in the absence of $T_c$ symmetry $(N=4,K=3.9,\kbar=1$). Experimental data (symbols) is very well described by Eq.~(\ref{eq:6}) with $t_{\rm loc}$ and $t_{\rm E}^*$ treated as fitting parameters, and show a time singularity at $t_{\rm E}^*=10.2$ (green full line). The data is incompatible with $t_E^*=0$ (pink dotted line). Inset: Same as main panel, showing $\delta \langle p^2(t) \rangle$ instead. The black dashed line represents the diffusion at early times.}
\label{fig:2}
\end{figure}

We proceed to study the periodically-shifted QKR with $N=4$, for which $T_c$ symmetry is broken. For each experiment,  a phase configuration $\{a(1),a(2),a(3),a(4)\}$ is picked randomly~\cite{Note3}. We choose $K=3.9$, for which we have checked numerically that the classical dynamics is strongly chaotic, and $\kbar=1$.  We repeat the experiment for $100$ phase configurations. The inset of Fig.~\ref{fig:2} displays the measurements of $\delta \langle p^2(t)\rangle$ averaged over the configurations.

To analyze the experimental data, we use the Ehrenfest time-dependent weak dynamical localization theory~\cite{Tian:EhrenfestTimeDynamicalLoc:PRB05} in the case of broken $T_c$ symmetry. We can generalize straightforwardly the previous result of Ref.~\cite{Tian:EhrenfestTimeDynamicalLoc:PRB05} to the present system, obtaining
\begin{eqnarray}\label{eq:6}
 \delta \langle p^2(t) \rangle=2 D_0 \left( t -\frac{1}{t_{\rm loc}}\theta(t-t_{\rm E1}^*)(t-t_{\rm E1}^*)^{2}\, \right.\nonumber\\
 \left.+\frac{1}{2t_{\rm loc}}\theta(t-t_{\rm E2}^*)(t-t_{\rm E2}^*)^{2} \right) .
\end{eqnarray}
The Ehrenfest times $t_{\rm E1,2}^*$ are different in general. Comparing Eqs.~(\ref{eq:4}) and (\ref{eq:6}), we find that the power $3/2$-law is replaced by a quadratic law. This is because when the $T_c$ symmetry is broken, the coherent returning to the initial state or its time reversal is via more complicated interference patterns [e.g., (ii) in Fig.~\ref{fig:3}]. Note that in Eq.~(\ref{eq:6}) the second term is negative, arising from the coherent returning to the time reversal of initial state, whereas the third term is positive, arising from the coherent returning to the initial state, and has the same origin as the coherent forward scattering in disordered systems, which has recently been observed experimentally in the QKR~\cite{Karpiuk:CFSFirst:PRL12,Hainaut:CFS:arXiv17}.
We use the same fitting procedure as above, to extract first $D_0$ and then $t_{\rm loc}$, $t_{\rm E1}^*$ and $t_{\rm E2}^*$. Excellent agreement with measurements is found if we set $t_{\rm E1}^*=t_{\rm E2}^*(\equiv t_{\rm E}^*)$
(Fig.~\ref{fig:2}), which yields $t_{\rm E}^{*}=10.2$ and $t_{\rm loc}= 163$. We also see that when $t_{\rm E}^*$ is set to zero, Eq.~(\ref{eq:6}) deviates from measurements. The measurements of $\delta \langle p^2(t) \rangle$ provide us strong evidence for a time singularity at finite $t_{\rm E}^*$, with:
\begin{equation}\label{eq:3}
    m=2,\quad \nu=0,
\end{equation} 
which signals a sudden change in system's memory behavior.

In conclusion, our measurements are fully compatible with the existence of a DQPT in chaotic systems, with the critical time being the Ehrenfest time. We took advantage of the flexibility of ultracold atoms experiments to finely study exotic effects in the short time dynamics of quantum chaotic systems. This paves the way for a detailed study of the all-time dynamics of such systems. Finally, we remark that the QKR is a one-body system. In the future, it would be interesting to theoretically investigate the time singularity in many-body quantum chaos, notably in the Sachdev-Ye-Kitaev model \cite{Kitaev15,Sachdev93:PRL} currently under intensive investigations, and the ensuing connections to the AdS/CFT duality \cite{Maldacena16:PRD}.

\begin{acknowledgments}
We thank V. Galitski, J. Wang and L. W. Zhou for useful discussions. This work is supported by Agence Nationale de la Recherche (Grant K-BEC No. ANR-13-BS04-0001-01), the Labex CEMPI (Grant No. ANR-11-LABX-0007-01), the Ministry of Higher Education and Research, Hauts de France Council and European Regional Development Fund (ERDF) through the Contrat de Projets Etat-Region (CPER Photonics for Society, P4S), and the National Natural Science Foundation of China (Grants No. 11535011 and No. 11647601).
\end{acknowledgments}


\clearpage

\renewcommand{\thesection}{S\arabic{section}}
\renewcommand{\thesubsection}{\thesection.\arabic{subsection}}
\renewcommand{\theequation}{S\arabic{equation}}
\renewcommand{\thefigure}{S\arabic{figure}}

\setcounter{page}{1}
\setcounter{equation}{0}
\setcounter{figure}{0}

\begin{widetext}

	\section{Supplemental Material for: Experimental observation of time singularity in classical-to-quantum chaos transition}
	
	\subsection{The origin of time singularity}
	
	The return probability, $L(t)=|\langle p|\hat U^t|p\rangle|^2$, for standard QKR has been shown \cite{Tian:EhrenfestTimeDynamicalLoc:PRB05,Tian:TheoryOfLocalizationQKR:NJP10} to be
	\begin{equation}\label{eq:S3}
	L(t)=\int\frac{d\omega}{2\pi}e^{-i\omega t}\int\frac{d\varphi}{2\pi}{\cal Y}(\varphi,\omega),
	\end{equation}
	where
	\begin{equation}\label{eq:S4}
	{\cal Y}(\varphi,\omega)=\frac{1}{-i\omega+D(\omega)\varphi^2}
	\end{equation}
	with $D(\omega)$ being the frequency-dependent diffusion coefficient. On the other hand, the cloud expansion
	\begin{equation}\label{eq:S5}
	\delta\langle p^2(t)\rangle=-\int\frac{d\omega}{2\pi}e^{-i\omega t}\partial^2_\varphi|_{\varphi \rightarrow 0}{\cal Y}(\varphi,\omega).
	\end{equation}
	Equations (\ref{eq:S3}), (\ref{eq:S4}) and (\ref{eq:S5}) show that $L(t)$ and $\delta\langle p^2(t)\rangle$ are mutually related. Specifically, Eq.~(5) is equivalent to \cite{Tian:EhrenfestTimeDynamicalLoc:PRB05,Tian:EhrenfestTimeDynamicalLoc:PRL04}
	\begin{equation}\label{eq:S6}
	D(\omega)=D_0-\frac{\kbar}{\pi}\int\frac{d\varphi e^{i\omega t_{\rm E}^*}}{-i\omega+D_0\varphi^2}.
	\end{equation}
	When substitute it into Eq.~(\ref{eq:S4}), we obtain
	\begin{equation}\label{eq:7}
	L(t)=\frac{1}{\sqrt{4\pi D_0 t}}+\frac{1}{4}\frac{\kbar}{D_0} \theta(t-t_{\rm E}^*),
	\end{equation}
	which is nonanalytic at $t=t_{\rm E}^*$.
	
	\subsection{The time correlation of angular position}
	
	We derive a relation between the time correlation of rotor's angular position $
	\hat x$, defined as \cite{Shepelyansky87:PhysicaD}
	\begin{equation}\label{eq:S7}
	C_q(t)\equiv \frac{K^2}{2}\langle 0|\sin\hat x(t)\sin\hat x(0)+\sin\hat x(0)\sin\hat x(t)|0\rangle,
	\end{equation}
	and the observable defined by Eq.~(3). To simplify discussions we focus on the standard QKR. The Green-Kubo formula for the static diffusion coefficient $D(\omega\rightarrow 0)$ was given in Ref.~\cite{Shepelyansky87:PhysicaD}, read
	\begin{equation}\label{eq:S8}
	D(\omega\rightarrow 0)=\int_{-\infty}^\infty dt C_q(t).
	\end{equation}
	This formula can be generalized straightforwardly to the dynamic diffusion coefficient $D(\omega)$ as
	\begin{equation}\label{eq:S9}
	D(\omega)=\int_{-\infty}^\infty dt e^{i\omega t}C_q(t).
	\end{equation}
	On the other hand, Eq.~(\ref{eq:S5}) gives
	\begin{equation}\label{eq:S10}
	\delta\langle p^2(t)\rangle=-\int\frac{d\omega}{\pi}e^{-i\omega t}\frac{D(\omega)}{\omega^2}.
	\end{equation}
	\begin{equation}\label{eq:S11}
	C_q(t)=\frac{1}{2}\frac{d^2}{dt^2}\delta\langle p^2(t)\rangle.
	\end{equation}
	
	\subsection{Spectral statistics of the periodically-shifted QKR}
	
	Because of $a(t+N)=a(t)$, the Floquet operator ${\hat U}$ of the periodically-shifted QKR is a product of $N$ one-step evolution operators, i.e.,
	\begin{equation}\label{eq:S1}
	{\hat U}=\prod_{j=-(N-1)}^{0}{\hat U}(j),\quad {\hat U}(j) \equiv e^{-\frac{i}{4\kbar} \hat p^2} e^{-i\frac{K}{\kbar} \cos(\hat x - a(j))} e^{-\frac{i}{4\kbar} \hat p^2}.
	\end{equation}
	The transposition (denoted as `${\rm T}$') of this operator is
	\begin{equation}\label{eq:S2}
	\hat U^{\rm T}=\prod_{j=0}^{N-1}\hat U^{\rm T}(-j)=\prod_{j=-(N-1)}^{0}\hat U^{\rm T}(-(j-(N-1))),\quad \hat U^{\rm T}(j) = e^{-\frac{i}{4\kbar} \hat p^2} e^{-i\frac{K}{\kbar} \cos(\hat x + a(j))} e^{-\frac{i}{4\kbar} \hat p^2}.
	\end{equation}
	The system has the $T_c$ symmetry if $\hat U^{\rm T} = \hat U$, up to unitary transformations corresponding to space and time translations (allowing to choose the axis of symmetry for both parity and time reversal transformations). For instance, the choice $a(j) = -a(N-1-j)$ satisfies this criterion.

	\begin{figure}[h]
		\center
		\includegraphics[width=16.0cm]{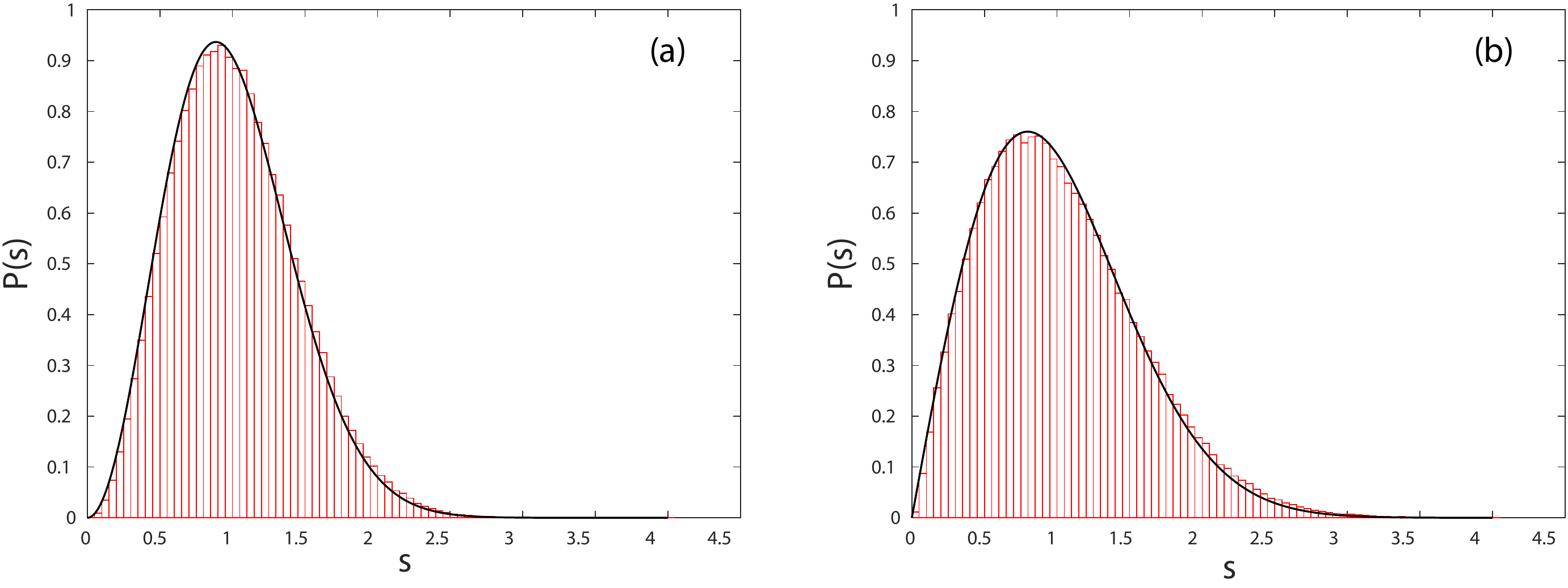}
		\caption{Simulations (histograms) show that the level spacing distribution $P(s)$ of the periodically-shifted QKR is in excellent agreement with the Wigner surmise (solid line) of orthogonal type, when the $T_c$ symmetry is present (a), and of unitary type when the symmetry is broken (b).}
		\label{fig:S1}
	\end{figure}
	
	We study numerically the spectral fluctuations of the periodically-shifted QKR with (broken) $T_c$ symmetry. We consider $N=4$. To realize a periodically-shifted QKR with $T_c$ symmetry, we choose the modulation configuration with $a(3)=-a(0)$ and $a(2)=-a(1)$, where $a(0)$ and $a(1)$ are sampled randomly, whereas to realize a periodically-shifted QKR with broken $T_c$ symmetry, the four phases are picked at random. We truncate in the momentum basis $p=m\kbar$, with a $p_{\rm max}$ small compared to the localization length. More precisely, we evolve all states $|p\rangle$ ($p=-p_{\rm max},\cdots,p_{\rm max}$) over one period using the standard fast Fourier transformation technique to reconstruct the matrix element $(\hat U)_{pp'}$. We then diagonalize the matrix $\{(\hat U)_{pp'}\}$ to obtain its eigenvalues $\{e^{-i \epsilon_n}\}$ and order the quasi-energies $\{\epsilon_n \in[0,2\pi[\}$. From this we compute the distribution of the normalized level spacing $s_n=\frac{\epsilon_{n+1}-\epsilon_n}{\delta}$, where $\delta$ is the mean level spacing. We randomly sample about $10000$ modulation configurations with phase configurations corresponding to the (broken) $T_c$ symmetry. The results are shown in Fig.~\ref{fig:S1}. We see that the distribution $P(s)$ for the periodically-shifted QKR with (broken) $T_c$ symmetry is in excellent agreement with the Wigner surmise for the orthogonal (unitary) class.


\begin{thebibliography}{36}
	\expandafter\ifx\csname natexlab\endcsname\relax\def\natexlab#1{#1}\fi
	\expandafter\ifx\csname bibnamefont\endcsname\relax
	\def\bibnamefont#1{#1}\fi
	\expandafter\ifx\csname bibfnamefont\endcsname\relax
	\def\bibfnamefont#1{#1}\fi
	\expandafter\ifx\csname citenamefont\endcsname\relax
	\def\citenamefont#1{#1}\fi
	\expandafter\ifx\csname url\endcsname\relax
	\def\url#1{\texttt{#1}}\fi
	\expandafter\ifx\csname urlprefix\endcsname\relax\def\urlprefix{URL }\fi
	\providecommand{\bibinfo}[2]{#2}
	\providecommand{\eprint}[2][]{\url{#2}}
	
	\bibitem[{\citenamefont{Hayden and Preskill}(2007)}]{Preskill07:JHEP}
	\bibinfo{author}{\bibfnamefont{P.}~\bibnamefont{Hayden}} \bibnamefont{and}
	\bibinfo{author}{\bibfnamefont{J.}~\bibnamefont{Preskill}},
	\bibinfo{journal}{J. High Energy Phys.} \textbf{\bibinfo{volume}{2007}},
	\bibinfo{pages}{120} (\bibinfo{year}{2007}).
	
	\bibitem[{\citenamefont{Kitaev}(2014)}]{Kitaev15}
	\bibinfo{author}{\bibfnamefont{A.}~\bibnamefont{Kitaev}},
	\bibinfo{journal}{{talk given at Fundamental Physics Prize Symposium}}
	(\bibinfo{year}{2014}).
	
	\bibitem[{\citenamefont{Kitaev}(2015)}]{Kitaev15:KITP}
	\bibinfo{author}{\bibfnamefont{A.}~\bibnamefont{Kitaev}},
	\bibinfo{journal}{{talks at KITP, April 7, 2015 and May 27, 2015}}
	(\bibinfo{year}{2015}).
	
	\bibitem[{\citenamefont{{Maldacena} and {Stanford}}(2016)}]{Maldacena16:PRD}
	\bibinfo{author}{\bibfnamefont{J.}~\bibnamefont{{Maldacena}}} \bibnamefont{and}
	\bibinfo{author}{\bibfnamefont{D.}~\bibnamefont{{Stanford}}},
	\bibinfo{journal}{Phys. Rev. D} \textbf{\bibinfo{volume}{94}},
	\bibinfo{pages}{106002} (\bibinfo{year}{2016}).
	
	\bibitem[{\citenamefont{{Maldacena} et~al.}(2016)\citenamefont{{Maldacena},
			{Shenker}, and {Stanford}}}]{Maldacena16:JHEP}
	\bibinfo{author}{\bibfnamefont{J.}~\bibnamefont{{Maldacena}}},
	\bibinfo{author}{\bibfnamefont{S.~H.} \bibnamefont{{Shenker}}},
	\bibnamefont{and}
	\bibinfo{author}{\bibfnamefont{D.}~\bibnamefont{{Stanford}}},
	\bibinfo{journal}{J. High Energy Phys.} \textbf{\bibinfo{volume}{2016}},
	\bibinfo{pages}{106} (\bibinfo{year}{2016}).
	
	\bibitem[{\citenamefont{{Patel} and {Sachdev}}(2017)}]{Sachdev17:PNAS}
	\bibinfo{author}{\bibfnamefont{A.~A.} \bibnamefont{{Patel}}} \bibnamefont{and}
	\bibinfo{author}{\bibfnamefont{S.}~\bibnamefont{{Sachdev}}},
	\bibinfo{journal}{Proc. Nat. Acad. Sci.} \textbf{\bibinfo{volume}{114}},
	\bibinfo{pages}{1844} (\bibinfo{year}{2017}).
	
	\bibitem[{\citenamefont{{Aleiner} et~al.}(2016)\citenamefont{{Aleiner},
			{Faolo}, and {Ioffe}}}]{Ioffe16:Ann.Phys.}
	\bibinfo{author}{\bibfnamefont{I.~L.} \bibnamefont{{Aleiner}}},
	\bibinfo{author}{\bibfnamefont{L.}~\bibnamefont{{Faolo}}}, \bibnamefont{and}
	\bibinfo{author}{\bibfnamefont{L.~B.} \bibnamefont{{Ioffe}}},
	\bibinfo{journal}{Ann. Phys. (N. Y.)} \textbf{\bibinfo{volume}{375}},
	\bibinfo{pages}{378} (\bibinfo{year}{2016}).
	
	\bibitem[{\citenamefont{Chen and
			Tian}(2014)}]{Chen:QuantumDrivenIntegerQuantumHallEffect:PRL14}
	\bibinfo{author}{\bibfnamefont{Y.}~\bibnamefont{Chen}} \bibnamefont{and}
	\bibinfo{author}{\bibfnamefont{C.}~\bibnamefont{Tian}},
	\bibinfo{journal}{Phys. Rev. Lett.} \textbf{\bibinfo{volume}{113}},
	\bibinfo{pages}{216802} (\bibinfo{year}{2014}).
	
	\bibitem[{\citenamefont{Tian et~al.}(2016)\citenamefont{Tian, Chen, and
			Wang}}]{Tian:EmergencerQuantumHallEffectChaos:PRB16}
	\bibinfo{author}{\bibfnamefont{C.}~\bibnamefont{Tian}},
	\bibinfo{author}{\bibfnamefont{Y.}~\bibnamefont{Chen}}, \bibnamefont{and}
	\bibinfo{author}{\bibfnamefont{J.}~\bibnamefont{Wang}},
	\bibinfo{journal}{Phys. Rev. B} \textbf{\bibinfo{volume}{93}},
	\bibinfo{pages}{075403} (\bibinfo{year}{2016}).
	
	\bibitem[{\citenamefont{Shepelyansky}(1987)}]{Shepelyansky87:PhysicaD}
	\bibinfo{author}{\bibfnamefont{D.~L.} \bibnamefont{Shepelyansky}},
	\bibinfo{journal}{Physica D} \textbf{\bibinfo{volume}{28}},
	\bibinfo{pages}{103} (\bibinfo{year}{1987}).
	
	\bibitem[{\citenamefont{Casati and Chirikov}(1995)}]{CasatiChirikov95}
	\bibinfo{author}{\bibfnamefont{G.}~\bibnamefont{Casati}} \bibnamefont{and}
	\bibinfo{author}{\bibfnamefont{B.~V.} \bibnamefont{Chirikov}},
	\emph{\bibinfo{title}{{Quantum Chaos: Between Order and Disorder}}}
	(\bibinfo{publisher}{{Cambridge University Press}},
	\bibinfo{address}{{Cambridge, UK}}, \bibinfo{year}{1995}).
	
	\bibitem[{\citenamefont{{Larkin} and {Aleiner}}(1996)}]{Larkin96:PRB}
	\bibinfo{author}{\bibfnamefont{I.~L.} \bibnamefont{{Aleiner}}} \bibnamefont{and}
	\bibinfo{author}{\bibfnamefont{A.~I.} \bibnamefont{{Larkin}}},
	\bibinfo{journal}{Phys. Rev. B} \textbf{\bibinfo{volume}{54}},
	\bibinfo{pages}{14423} (\bibinfo{year}{1996}).
	
	\bibitem[{\citenamefont{{Sokolov} and {Zhirov}}(2008)}]{Sokolov08:EPL}
	\bibinfo{author}{\bibfnamefont{V.~V.} \bibnamefont{{Sokolov}}}
	\bibnamefont{and} \bibinfo{author}{\bibfnamefont{O.~V.}
		\bibnamefont{{Zhirov}}}, \bibinfo{journal}{EPL}
	\textbf{\bibinfo{volume}{84}}, \bibinfo{pages}{3001} (\bibinfo{year}{2008}).
	
	\bibitem[{\citenamefont{{Benenti} and {Casati}}(2009)}]{Casati09:PRE}
	\bibinfo{author}{\bibfnamefont{G.}~\bibnamefont{{Benenti}}} \bibnamefont{and}
	\bibinfo{author}{\bibfnamefont{G.}~\bibnamefont{{Casati}}},
	\bibinfo{journal}{Phys. Rev. E} \textbf{\bibinfo{volume}{79}},
	\bibinfo{pages}{025201(R)} (\bibinfo{year}{2009}).
	
	\bibitem[{\citenamefont{Casati et~al.}(1979)\citenamefont{Casati, Chirikov,
			Ford, and Izrailev}}]{Casati:LocDynFirst:LNP79}
	\bibinfo{author}{\bibfnamefont{G.}~\bibnamefont{Casati}},
	\bibinfo{author}{\bibfnamefont{B.~V.} \bibnamefont{Chirikov}},
	\bibinfo{author}{\bibfnamefont{J.}~\bibnamefont{Ford}}, \bibnamefont{and}
	\bibinfo{author}{\bibfnamefont{F.~M.} \bibnamefont{Izrailev}},
	\bibinfo{journal}{Lecture Notes in Physics} \textbf{\bibinfo{volume}{93}},
	\bibinfo{pages}{334} (\bibinfo{year}{1979}).
	
	\bibitem[{\citenamefont{{Larkin} and
			{Ovchinnikov}}(1969)}]{Larkin:Ehrenfest_time:JETP68}
	\bibinfo{author}{\bibfnamefont{A.~I.} \bibnamefont{{Larkin}}} \bibnamefont{and}
	\bibinfo{author}{\bibfnamefont{Y.~N.} \bibnamefont{{Ovchinnikov}}},
	\bibinfo{journal}{Sov. J. Exp. Theor. Phys.} \textbf{\bibinfo{volume}{28}},
	\bibinfo{pages}{1200} (\bibinfo{year}{1969}).
	
	\bibitem[{\citenamefont{Tian et~al.}(2004)\citenamefont{Tian, Kamenev, and
			Larkin}}]{Tian:EhrenfestTimeDynamicalLoc:PRL04}
	\bibinfo{author}{\bibfnamefont{C.}~\bibnamefont{Tian}},
	\bibinfo{author}{\bibfnamefont{A.}~\bibnamefont{Kamenev}}, \bibnamefont{and}
	\bibinfo{author}{\bibfnamefont{A.}~\bibnamefont{Larkin}},
	\bibinfo{journal}{Phys. Rev. Lett.} \textbf{\bibinfo{volume}{93}},
	\bibinfo{pages}{124101} (\bibinfo{year}{2004}).
	
	\bibitem[{\citenamefont{Tian et~al.}(2005)\citenamefont{Tian, Kamenev, and
			Larkin}}]{Tian:EhrenfestTimeDynamicalLoc:PRB05}
	\bibinfo{author}{\bibfnamefont{C.}~\bibnamefont{Tian}},
	\bibinfo{author}{\bibfnamefont{A.}~\bibnamefont{Kamenev}}, \bibnamefont{and}
	\bibinfo{author}{\bibfnamefont{A.}~\bibnamefont{Larkin}},
	\bibinfo{journal}{Phys. Rev. B} \textbf{\bibinfo{volume}{72}},
	\bibinfo{pages}{045108} (\bibinfo{year}{2005}).
	
	\bibitem[{\citenamefont{Rozenbaum et~al.}(2017)\citenamefont{Rozenbaum,
			Ganeshan, and Galitski}}]{Galitskii:out_of_time_order_correlator}
	\bibinfo{author}{\bibfnamefont{E.~B.} \bibnamefont{Rozenbaum}},
	\bibinfo{author}{\bibfnamefont{S.}~\bibnamefont{Ganeshan}}, \bibnamefont{and}
	\bibinfo{author}{\bibfnamefont{V.}~\bibnamefont{Galitski}},
	\bibinfo{journal}{Phys. Rev. Lett.} \textbf{\bibinfo{volume}{118}},
	\bibinfo{pages}{086801} (\bibinfo{year}{2017}).
	
	\bibitem[{SM()}]{SM}
	\bibinfo{note}{See Supplemental Material}.
	
	\bibitem[{\citenamefont{Heyl et~al.}(2013)\citenamefont{Heyl, Polkovnikov, and
			Kehrein}}]{Heyl:DynamicalPhaseTransition:PRL13}
	\bibinfo{author}{\bibfnamefont{M.}~\bibnamefont{Heyl}},
	\bibinfo{author}{\bibfnamefont{A.}~\bibnamefont{Polkovnikov}},
	\bibnamefont{and} \bibinfo{author}{\bibfnamefont{S.}~\bibnamefont{Kehrein}},
	\bibinfo{journal}{Phys. Rev. Lett.} \textbf{\bibinfo{volume}{110}},
	\bibinfo{pages}{135704} (\bibinfo{year}{2013}).
	
	\bibitem[{\citenamefont{Jurcevic et~al.}(2017)\citenamefont{Jurcevic, Shen,
			Hauke, Maier, Brydges, Hempel, Lanyon, Heyl, Blatt, and
			Roos}}]{Jurcevic:DynamicalPhaseTransition:PRL17}
	\bibinfo{author}{\bibfnamefont{P.}~\bibnamefont{Jurcevic}},
	\bibinfo{author}{\bibfnamefont{H.}~\bibnamefont{Shen}},
	\bibinfo{author}{\bibfnamefont{P.}~\bibnamefont{Hauke}},
	\bibinfo{author}{\bibfnamefont{C.}~\bibnamefont{Maier}},
	\bibinfo{author}{\bibfnamefont{T.}~\bibnamefont{Brydges}},
	\bibinfo{author}{\bibfnamefont{C.}~\bibnamefont{Hempel}},
	\bibinfo{author}{\bibfnamefont{B.~P.} \bibnamefont{Lanyon}},
	\bibinfo{author}{\bibfnamefont{M.}~\bibnamefont{Heyl}},
	\bibinfo{author}{\bibfnamefont{R.}~\bibnamefont{Blatt}}, \bibnamefont{and}
	\bibinfo{author}{\bibfnamefont{C.~F.} \bibnamefont{Roos}},
	\bibinfo{journal}{Phys. Rev. Lett.} \textbf{\bibinfo{volume}{119}},
	\bibinfo{pages}{080501} (\bibinfo{year}{2017}).
	
	\bibitem[{\citenamefont{Heyl}(2017)}]{Heyl:DynamicalPhaseTransition:arXiv17}
	\bibinfo{author}{\bibfnamefont{M.}~\bibnamefont{Heyl}} (\bibinfo{year}{2017}),
	\eprint{1709.07461}.
	
	\bibitem[{\citenamefont{Hainaut
			et~al.}(2017{\natexlab{a}})\citenamefont{Hainaut, Manai, Chicireanu,
			Cl\'ement, Zemmouri, Garreau, Szriftgiser, Lemari\'e, Cherroret, and
			Delande}}]{Hainaut:ERO:PRL17}
	\bibinfo{author}{\bibfnamefont{C.}~\bibnamefont{Hainaut}},
	\bibinfo{author}{\bibfnamefont{I.}~\bibnamefont{Manai}},
	\bibinfo{author}{\bibfnamefont{R.}~\bibnamefont{Chicireanu}},
	\bibinfo{author}{\bibfnamefont{J.-F.} \bibnamefont{Cl\'ement}},
	\bibinfo{author}{\bibfnamefont{S.}~\bibnamefont{Zemmouri}},
	\bibinfo{author}{\bibfnamefont{J.C.}~\bibnamefont{Garreau}},
	\bibinfo{author}{\bibfnamefont{P.}~\bibnamefont{Szriftgiser}},
	\bibinfo{author}{\bibfnamefont{G.}~\bibnamefont{Lemari\'e}},
	\bibinfo{author}{\bibfnamefont{N.}~\bibnamefont{Cherroret}},
	\bibnamefont{and} \bibinfo{author}{\bibfnamefont{D.}~\bibnamefont{Delande}},
	\bibinfo{journal}{Phys. Rev. Lett.} \textbf{\bibinfo{volume}{118}},
	\bibinfo{pages}{184101} (\bibinfo{year}{2017}{\natexlab{a}}).
	
	\bibitem[{\citenamefont{Manai et~al.}(2015)\citenamefont{Manai, Cl{\'e}ment,
			Chicireanu, Hainaut, Garreau, Szriftgiser, and
			Delande}}]{Manai:Anderson2DKR:PRL15}
	\bibinfo{author}{\bibfnamefont{I.}~\bibnamefont{Manai}},
	\bibinfo{author}{\bibfnamefont{J.-F.} \bibnamefont{Cl{\'e}ment}},
	\bibinfo{author}{\bibfnamefont{R.}~\bibnamefont{Chicireanu}},
	\bibinfo{author}{\bibfnamefont{C.}~\bibnamefont{Hainaut}},
	\bibinfo{author}{\bibfnamefont{J.~C.} \bibnamefont{Garreau}},
	\bibinfo{author}{\bibfnamefont{P.}~\bibnamefont{Szriftgiser}},
	\bibnamefont{and} \bibinfo{author}{\bibfnamefont{D.}~\bibnamefont{Delande}},
	\bibinfo{journal}{Phys. Rev. Lett.} \textbf{\bibinfo{volume}{115}},
	\bibinfo{pages}{240603} (\bibinfo{year}{2015}).
	
	\bibitem [{Note1()}]{Note1}%
	\BibitemOpen
	\bibinfo {note} {In order to determine \unexpanded{$\delta \langle p^2(t) \rangle $}, we use a two-component test function~\cite{LobkisWeaver:SelfConsistentTransportLocWaves:PRE05} to fit the shape of measured $p^2 \Pi (p,t)$. We verified that this procedure agrees very well with results of numerical simulations.}\BibitemShut {Stop}%
	
	\bibitem[{\citenamefont{Fishman et~al.}(1982)\citenamefont{Fishman, Grempel,
			and Prange}}]{Fishman:LocDynAnders:PRL82}
	\bibinfo{author}{\bibfnamefont{S.}~\bibnamefont{Fishman}},
	\bibinfo{author}{\bibfnamefont{D.~R.} \bibnamefont{Grempel}},
	\bibnamefont{and} \bibinfo{author}{\bibfnamefont{R.~E.}
		\bibnamefont{Prange}}, \bibinfo{journal}{Phys. Rev. Lett.}
	\textbf{\bibinfo{volume}{49}}, \bibinfo{pages}{509} (\bibinfo{year}{1982}).
	
	
	\bibitem [{Note2()}]{Note2}%
	\BibitemOpen
	\bibinfo {note} {To consider the kicks as Dirac functions, the distance traveled by the atoms during the kicks should be much smaller than the wavelength of the standing-wave. This enforces $\tau\sim 200\, {\rm ns}$ for the present experimental setup. On the other hand, the condition: $\tau\ll T_1$ must be met also. Putting these two constraints together, we require $\kbar \geq 1$.}\BibitemShut {Stop}%
	
	
	\bibitem[{\citenamefont{Casati et~al.}(1989)\citenamefont{Casati, Guarneri, and
			Shepelyansky}}]{Casati:IncommFreqsQKR:PRL89}
	\bibinfo{author}{\bibfnamefont{G.}~\bibnamefont{Casati}},
	\bibinfo{author}{\bibfnamefont{I.}~\bibnamefont{Guarneri}}, \bibnamefont{and}
	\bibinfo{author}{\bibfnamefont{D.~L.} \bibnamefont{Shepelyansky}},
	\bibinfo{journal}{Phys. Rev. Lett.} \textbf{\bibinfo{volume}{62}},
	\bibinfo{pages}{345} (\bibinfo{year}{1989}).
	
	\bibitem[{\citenamefont{Hainaut
			et~al.}(2017{\natexlab{b}})\citenamefont{Hainaut, Manai, Cl\'ement, Garreau,
			Szriftgiser, Lemari\'e, Cherroret, Delande, and
			Chicireanu}}]{Hainaut:CFS:arXiv17}
	\bibinfo{author}{\bibfnamefont{C.}~\bibnamefont{Hainaut}},
	\bibinfo{author}{\bibfnamefont{I.}~\bibnamefont{Manai}},
	\bibinfo{author}{\bibfnamefont{J.-F.} \bibnamefont{Cl\'ement}},
	\bibinfo{author}{\bibfnamefont{J.}~\bibnamefont{Garreau}},
	\bibinfo{author}{\bibfnamefont{P.}~\bibnamefont{Szriftgiser}},
	\bibinfo{author}{\bibfnamefont{G.}~\bibnamefont{Lemari\'e}},
	\bibinfo{author}{\bibfnamefont{N.}~\bibnamefont{Cherroret}},
	\bibinfo{author}{\bibfnamefont{D.}~\bibnamefont{Delande}}, \bibnamefont{and}
	\bibinfo{author}{\bibfnamefont{R.}~\bibnamefont{Chicireanu}}
	(\bibinfo{year}{2017}{\natexlab{b}}), \eprint{{1709.02632}}.
	
	\bibitem[{\citenamefont{Efetov}(1997)}]{Efetov:SupersymmetryInDisorder:97}
	\bibinfo{author}{\bibfnamefont{K.}~\bibnamefont{Efetov}},
	\emph{\bibinfo{title}{{Supersymmetry in Disorder and Chaos}}}
	(\bibinfo{publisher}{{Cambridge University Press}},
	\bibinfo{address}{{Cambridge, UK}}, \bibinfo{year}{1997}).
	
	\bibitem[{\citenamefont{Tian and
			Altland}(2010)}]{Tian:TheoryOfLocalizationQKR:NJP10}
	\bibinfo{author}{\bibfnamefont{C.}~\bibnamefont{Tian}} \bibnamefont{and}
	\bibinfo{author}{\bibfnamefont{A.}~\bibnamefont{Altland}},
	\bibinfo{journal}{New J. Phys} \textbf{\bibinfo{volume}{12}},
	\bibinfo{pages}{043043} (\bibinfo{year}{2010}).
	
	\bibitem [{Note3()}]{Note3}%
	\BibitemOpen
	\bibinfo {note} {In principle antisymmetric sequences should be excluded (see Supplemental Materials \cite{SM}), but the probability of randomly picking one of them is negligible.}\BibitemShut {Stop}%
	
	\bibitem[{\citenamefont{Chirikov}(1979)}]{Chirikov:ChaosClassKR:PhysRep79}
	\bibinfo{author}{\bibfnamefont{B.~V.} \bibnamefont{Chirikov}},
	\bibinfo{journal}{Phys. Rep.} \textbf{\bibinfo{volume}{52}},
	\bibinfo{pages}{263} (\bibinfo{year}{1979}).
	
	\bibitem[{\citenamefont{Altland}(1993)}]{Altland:DiagrammaticAndersonLocQKR:PRL93}
	\bibinfo{author}{\bibfnamefont{A.}~\bibnamefont{Altland}},
	\bibinfo{journal}{Phys. Rev. Lett.} \textbf{\bibinfo{volume}{71}},
	\bibinfo{pages}{69} (\bibinfo{year}{1993}).
	
	\bibitem[{\citenamefont{{Gor'kov} et~al.}(1979)\citenamefont{{Gor'kov},
			{Larkin}, and {Khmel'nitskii}}}]{Gorkov:WeakLocalization:JETP79}
	\bibinfo{author}{\bibfnamefont{L.~P.} \bibnamefont{{Gor'kov}}},
	\bibinfo{author}{\bibfnamefont{A.~I.} \bibnamefont{{Larkin}}},
	\bibnamefont{and} \bibinfo{author}{\bibfnamefont{D.~E.}
		\bibnamefont{{Khmel'nitskii}}}, \bibinfo{journal}{J. Exp. Theor. Phys. Lett.}
	\textbf{\bibinfo{volume}{30}}, \bibinfo{pages}{228} (\bibinfo{year}{1979}).
	
	\bibitem[{\citenamefont{Karpiuk et~al.}(2012)\citenamefont{Karpiuk, Cherroret,
			Lee, Gr{\'e}maud, M{\"u}ller, and Miniatura}}]{Karpiuk:CFSFirst:PRL12}
	\bibinfo{author}{\bibfnamefont{T.}~\bibnamefont{Karpiuk}},
	\bibinfo{author}{\bibfnamefont{N.}~\bibnamefont{Cherroret}},
	\bibinfo{author}{\bibfnamefont{K.~L.} \bibnamefont{Lee}},
	\bibinfo{author}{\bibfnamefont{B.}~\bibnamefont{Gr{\'e}maud}},
	\bibinfo{author}{\bibfnamefont{C.~A.} \bibnamefont{M{\"u}ller}},
	\bibnamefont{and}
	\bibinfo{author}{\bibfnamefont{C.}~\bibnamefont{Miniatura}},
	\bibinfo{journal}{Phys. Rev. Lett.} \textbf{\bibinfo{volume}{109}},
	\bibinfo{pages}{190601} (\bibinfo{year}{2012}).
	
	\bibitem[{\citenamefont{{Sachdev} and {Ye}}(1993)}]{Sachdev93:PRL}
	\bibinfo{author}{\bibfnamefont{S.}~\bibnamefont{{Sachdev}}} \bibnamefont{and}
	\bibinfo{author}{\bibfnamefont{J.}~\bibnamefont{{Ye}}},
	\bibinfo{journal}{Phys. Rev. Lett.} \textbf{\bibinfo{volume}{70}},
	\bibinfo{pages}{3339} (\bibinfo{year}{1993}).
	
	\bibitem[{\citenamefont{Lobkis and
			Weaver}(2005)}]{LobkisWeaver:SelfConsistentTransportLocWaves:PRE05}
	\bibinfo{author}{\bibfnamefont{O.~I.} \bibnamefont{Lobkis}} \bibnamefont{and}
	\bibinfo{author}{\bibfnamefont{R.~L.} \bibnamefont{Weaver}},
	\bibinfo{journal}{Phys. Rev. E} \textbf{\bibinfo{volume}{71}},
	\bibinfo{pages}{011112} (\bibinfo{year}{2005}).
	
\end{thebibliography}

\begin{thebibliography}{4}
		\expandafter\ifx\csname natexlab\endcsname\relax\def\natexlab#1{#1}\fi
		\expandafter\ifx\csname bibnamefont\endcsname\relax
		\def\bibnamefont#1{#1}\fi
		\expandafter\ifx\csname bibfnamefont\endcsname\relax
		\def\bibfnamefont#1{#1}\fi
		\expandafter\ifx\csname citenamefont\endcsname\relax
		\def\citenamefont#1{#1}\fi
		\expandafter\ifx\csname url\endcsname\relax
		\def\url#1{\texttt{#1}}\fi
		\expandafter\ifx\csname urlprefix\endcsname\relax\def\urlprefix{URL }\fi
		\providecommand{\bibinfo}[2]{#2}
		\providecommand{\eprint}[2][]{\url{#2}}
		
		\bibitem[{\citenamefont{Tian et~al.}(2005)\citenamefont{Tian, Kamenev, and
				Larkin}}]{Tian:EhrenfestTimeDynamicalLoc:PRB05}
		\bibinfo{author}{\bibfnamefont{C.}~\bibnamefont{Tian}},
		\bibinfo{author}{\bibfnamefont{A.}~\bibnamefont{Kamenev}}, \bibnamefont{and}
		\bibinfo{author}{\bibfnamefont{A.}~\bibnamefont{Larkin}},
		\bibinfo{journal}{Phys. Rev. B} \textbf{\bibinfo{volume}{72}},
		\bibinfo{pages}{045108} (\bibinfo{year}{2005}).
		
		\bibitem[{\citenamefont{Tian and
				Altland}(2010)}]{Tian:TheoryOfLocalizationQKR:NJP10}
		\bibinfo{author}{\bibfnamefont{C.}~\bibnamefont{Tian}} \bibnamefont{and}
		\bibinfo{author}{\bibfnamefont{A.}~\bibnamefont{Altland}},
		\bibinfo{journal}{New J. Phys} \textbf{\bibinfo{volume}{12}},
		\bibinfo{pages}{043043} (\bibinfo{year}{2010}).
		
		\bibitem[{\citenamefont{Tian et~al.}(2004)\citenamefont{Tian, Kamenev, and
				Larkin}}]{Tian:EhrenfestTimeDynamicalLoc:PRL04}
		\bibinfo{author}{\bibfnamefont{C.}~\bibnamefont{Tian}},
		\bibinfo{author}{\bibfnamefont{A.}~\bibnamefont{Kamenev}}, \bibnamefont{and}
		\bibinfo{author}{\bibfnamefont{A.}~\bibnamefont{Larkin}},
		\bibinfo{journal}{Phys. Rev. Lett.} \textbf{\bibinfo{volume}{93}},
		\bibinfo{pages}{124101} (\bibinfo{year}{2004}).
		
		\bibitem[{\citenamefont{Shepelyansky}(1987)}]{Shepelyansky87:PhysicaD}
		\bibinfo{author}{\bibfnamefont{D.~L.} \bibnamefont{Shepelyansky}},
		\bibinfo{journal}{Physica D} \textbf{\bibinfo{volume}{28}},
		\bibinfo{pages}{103} (\bibinfo{year}{1987}).
		
	\end{thebibliography}

	\clearpage

\end{widetext}

\end{document}